\documentclass[%
 reprint,
 amsmath,amssymb,
 aps,
 prapplied,
]{revtex4-1}

\usepackage{float}
\usepackage{graphicx}%
\usepackage{dcolumn}%
\usepackage{bm}%
\usepackage{xcolor}
\usepackage[normalem]{ulem}
\usepackage{siunitx}

\usepackage{appendix}

\newcommand{\trm}[1]{\textrm{#1}}
\newcommand{\prens}[1]{\left(#1\right)}

\newcommand{\um}{\SI{}{\micro\meter}}
\newcommand{\nm}{\SI{}{\nano\meter}}
\newcommand{\LN}{$\text{LiNbO}_3$}

\definecolor{cjsblue}{HTML}{4b50a8}

\definecolor{slate}{HTML}{708090}

\begin{document}
\title{The role of resonance and bandgaps in high $k_\textrm{eff}^2$ transducers}

\author{Christopher J. Sarabalis}
\email{sicamor@stanford.edu}
\author{Yanni D. Dahmani}
\author{Agnetta Y. Cleland}
\author{Amir H. Safavi-Naeini}
\email{safavi@stanford.edu}

\affiliation{%
 Department of Applied Physics and Ginzton Laboratory, Stanford University\\
 348 Via Pueblo Mall, Stanford, California 94305, USA
}%

\date{\today}%

\begin{abstract}

Bandgaps formed in a piezoelectric transducer with large coupling, $k_\trm{eff}^2$, qualitatively modify its electrical response.  This regime in which electrical loading strongly couples forward and backward waves occurs in thin-film lithium niobate which has recently become available and amenable to nanopatterning.
In this work, we study how resonance and bandgaps modify the design and performance of transducers and delay lines in thin-film lithium niobate. These films are an attractive platform for GHz frequency applications in low-power RF analog signal processing, optomechanics, and quantum devices due to their high coupling, low loss, excellent optical properties, and compatibility with superconducting quantum circuits.  We demonstrate aluminum IDTs in this platform for horizontal shear (SH) waves between $1.2$ and $3.3$~GHz and longitudinal waves between $2.1$ and $5.4$~GHz.  For the SH waves, we measure a piezoelectric coupling coefficient of $13\%$ and $6.0$~dB/mm propagation losses in delay lines up to $1.2$~mm with a $300$~ns delay in air at room temperature.  Reflections from electrical loading when $k_\trm{eff}^2$ is large lead to a departure from the impulse response model widely used to model surface acoustic wave devices.  Finite element method models and an experimental finger-pair sweep are used to characterize the role of resonance in these transducers, illuminating the physics behind the anomalously large motional admittances of these small-footprint IDTs.

\end{abstract}

\maketitle

\section{Introduction}

Piezoelectric devices enjoy a wide range of modern applications in microwave, optical, and quantum systems thanks to the development of low-loss piezoelectric materials with large piezoelectric coupling such as lithium niobate~\cite{Morgan2002}.
The size, weight and power (SWaP) requirements of ultra-low power RF systems and the Internet of Things has renewed interest in piezoelectric analog signal processing~\cite{Olsson2014,Song2016,Manzaneque2017,Gong2017}. Parallel developments in photonic and optomechanical devices~\cite{Sarabalis2018,VanLaer2019,Khan2018,Kittlaus2015,Sohn2018,Tadesse2014} and efforts to improve the connectivity and scalability of emerging quantum hardware~\cite{Satzinger2018QuantumPhonons, Pechal2018SuperconductingStorage} has led to a fruitful convergence of several strands of research and brought new classes of acoustic structures to the forefront. The valuable library of techniques and intuitions developed for the design of surface acoustic wave devices can only be applied approximately in this new context. Improving our understanding of and ability to control the propagation and transduction of mechanical waves in these thin-film, highly coupled nanostructures is paramount for continued progress in the field.  

In this work we study acoustic wave transduction and propagation in thin-film lithium niobate. The piezoelectric coupling coefficient \(k_\trm{eff}^2\) and quality factor \(Q\) are essential figures of merit across applications. Large \(k_\trm{eff}^2\) is necessary for large fractional bandwidth microwave filters and acousto-optic modulators; large field of view acousto-optic deflectors; and small transducers for wavelength-scale waveguides and resonators.
Resonators supporting longitudinal and horizontal shear (SH) waves in piezoelectric slabs have been demonstrated with \(k_\trm{eff}^2\) as high as 30\% and quality factors above 1000 with frequencies between $100$ and $500~\text{MHz}$~\cite{Zaitsev2001,Olsson2014,Song2016,Pop2017}.
These waves which propagate in the plane of the film have also been used to make delay lines and chirped pulse compressors~\cite{Vidal-alvarez2017,Manzaneque2017}.
More recently in thin-film lithium niobate, waves propagating normal to the plane of the film have been used to make resonators supporting a $30~\trm{GHz}$ mode with a \(k_\trm{eff}^2\) of $1$\% and a \(Q\) over $300$.
We extend work on the SH and longitudinal waves to higher frequencies for applications in photonic circuits where photon interactions are mediated by phonons with wavelengths on the order of a micron and quantum acoustic systems with qubits operating above $2~\text{GHz}$. To better understand these transducers, we map out their band structure by varying the interdigitated transducer (IDT) pitch \(a\). We show that SH and longitudinal waves can be efficiently transduced from \(1.2\) to \(3.3~\trm{GHz}\) and \(2.1\) to \(5.4~\trm{GHz}\), respectively.
We measure a group velocity of \(4000~\trm{m/s}\) and attenuation constant of \(6.0~\trm{dB/mm}\)  for SH waves at $2$~GHz and use these waves to realize a \(1.2~\trm{mm}\) line with a \(300~\trm{ns}\) microwave delay.

More broadly, we develop a better understanding of these transducers in terms of their band structure and, by sweeping the number of finger-pairs \(N\), detail the role that resonances play in their response.
We numerically study how the conductance spectrum \(G\prens{\omega}\) varies going from the low \(k_\trm{eff}^2\) regime in which the impulse response model holds to the observed high \(k_\trm{eff}^2\) strongly-coupled regime where resonances dominate the response.
These resonances decrease the bandwidth and increase the peak conductance of the main lobe, accounting for the large motional admittances exhibited by these small-footprint IDTs.

\begin{figure}[h!]
    \centering
    \includegraphics{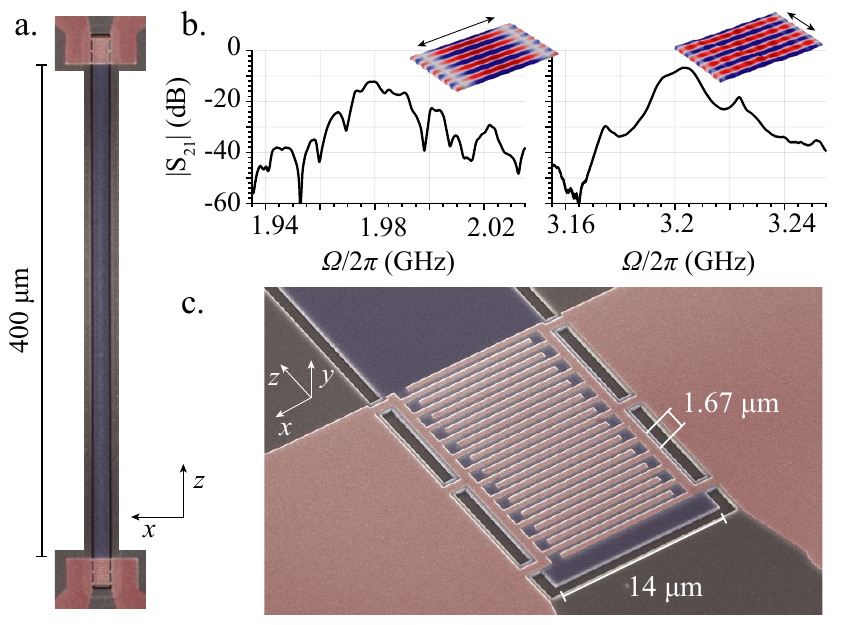}
    \caption{Straight-finger, uniform pitch and weight aluminum IDT's (red) are deposited on a 250~nm thick \LN~slab (blue) on silicon which is then released. \textbf{b.} The IDT's strongly transduce the 2~GHz SH modes (N~=~12) and the 3.2~GHz longitudinal response (N~=~16).}
    \label{fig:device}
\end{figure}

\section{S band transducers in suspended \LN}
\label{sec:S-band-IDTs}

We start by demonstrating efficient transduction of low-loss SH waves and longitudinal waves from $1.2$ to $5.4$~GHz before turning our attention to the physics of these high \(k_\trm{eff}^2\) IDTs in Section~\ref{sec:modeling-the-conductance}.

The delay lines studied here are $14~\um$ wide, $250~\nm$ thick plates of X-cut lithium niobate terminated on each end with aluminum, straight-finger IDTs with a $10~\um$ finger overlap.
The Y crystal axis is parallel and the extraordinary axis (Z) is perpendicular to the direction of propagation~\cite{Weis1985}.
The delay lines are tethered only in the transducer region, with six tethers at each transducer as shown in Figure~\ref{fig:device}.
Devices as long as $1.2~\trm{mm}$ were suspended in this manner without collapse. Transducers are etched along the back edge and therefore emit in one direction.

The fabrication methods are adapted from Arrangoiz-Arriola~\emph{et al.}~\cite{Arrangoiz-Arriola2018}.
Starting with wafers of $500~\nm$ X-cut \LN~on $500~\um$ of silicon, chips are diced and ion milled to $250~\nm$ before patterning the tethered membranes by electron beam lithography.
These patterns are argon ion-milled into the \LN, the mask is stripped with a piranha clean, and $100~\nm$ of aluminum is evaporated onto a PMMA 950 on PMMA 495 kilodalton lift-off bilayer in which electrodes were defined by e-beam lithography.
By correcting for proximity effect, IDTs with pitches ranging from $1~\um$ ($250~\nm$ fingers with $250~\nm$ gaps) to $2.2~\um$ could be made alongside contact pads with a single mask. After lift-off, the delay lines are released by underetching the silicon with XeF$_2$ vapor.

The S-parameters of the devices are measured with a vector network analyzer (Rhode \& Schwarz ZNB20 VNA) on a probe station with GSG probes (GGB nickel 40A). The probe station is calibrated with a GGB Industries CS-5 substrate to move the reference plane to the tips of the probes.
The magnitude of \(S_{21}(\omega)\) of 12 and 16 finger-pair IDT's separated by 100~\(\um\) is plotted in Figure~\ref{fig:device}b in which we see responses at 2 and \SI{3.2}{\giga\hertz} corresponding to the horizontal shear (SH) and longitudinal modes.
We calculate the single-port admittance $Y(\omega)$ from the $S(\omega)$ matrix~\cite{Pozar2009}.
As the conductance $G(\omega) \equiv \trm{Re}Y(\omega)$, and the susceptance $\chi(\omega) \equiv \trm{Im}Y(\omega)$ are a Hilbert transform pair, it is sufficient to restrict our attention to the conductance and the electrostatic capacitance \(C_\trm{s}\equiv-\lim_{\omega\rightarrow0} \chi_{11}(\omega)/\omega\).
For a brief overview of linear response theory and details on conventions used here, see Appendix~\ref{app:linear-response}.
In Figure~\ref{fig:bands}a, measurements of $G\prens{\omega}$ of an $N = 16$, $a = \SI{1.72}{\micro\meter}$ transducer for the SH band can be compared to results of a 2D finite element method (FEM) model discussed in detail in Section~\ref{sec:modeling-the-conductance}.

\begin{figure}
	\centering
	\includegraphics[width=0.85\linewidth]{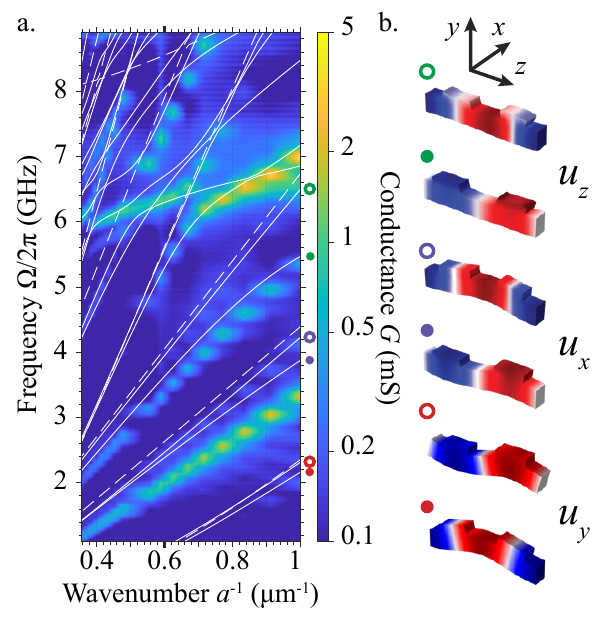}
	\caption{\textbf{a.} The band structure of the IDT is revealed in measurements of the conductance \(G\prens{\omega}\) by varying the pitch $a$.  Large peak conductance is measured between $1$ and $5$~GHz. \textbf{b.} A \SI{200}{\nano\meter} cross-section of the displacement \(u\) of the modes of the IDT at the X-point of an $a^{-1} = 0.6~\um^{-1}$ unit cell are shown at right for the Lamb, horizontal shear, and longitudinal bands. Waves propagate along $z$. The offset between the bands and the response is analyzed in Section~\ref{sec:modeling-the-conductance}.}
	\label{fig:pitch-sweep}
\end{figure}

\begin{figure*}
    \centering
    \includegraphics[width=0.7\linewidth]{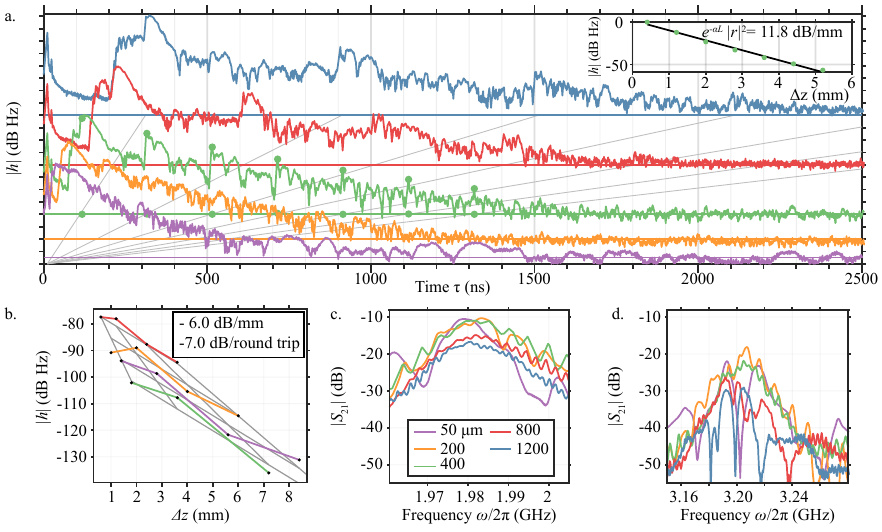}
    \caption{\textbf{a.} The impulse response \(h\prens{t}\) is plotted for delay lines of varied length.  For the \(L = 400~\um\) long line (green), the peak height \(h\prens{\tau_n}\) of the SH echoes are fit (inset) to infer  the round-trip loss \(r^2 e^{-\alpha L}\). \textbf{b.}  Peaks from the longest three lines are fit against the number of round-trips \(n\) and total propagation length \(\Delta z\) to find the propagation loss \(\alpha\) for the SH waves. In \textbf{c.} and \textbf{d.} we plot \(\left|S_{21}\right|\) for the SH and longitudinal modes.}
    \label{fig:L-sweep}
\end{figure*}

There are three sets of devices reported here in which the pitch, delay line length, and number of finger-pairs are swept.

The first set of devices consists of 19 IDTs with ten finger-pairs and a pitch \(a\) that increases uniformly from \(1\) to \(2.8~\um\). 
The pitch $a$ of the IDT determines the wavelength of the waves it excites.
Sweeping the pitch maps out the efficiently transduced IDT bands, apparent in the measured \(G(\omega)\) plotted in Figure~\ref{fig:pitch-sweep}a.
The band structure of a free slab (dashed) and the X-point frequencies of the IDT (solid) are computed by the FEM and overlaid on the measured conductance for comparison. 
The fundamental horizontal shear and longitudinal modes are seen to vary from 1.2 to 5.4 GHz, spanning the S and entering the C band.
We plot the displacement profile of these modes in Figure~\ref{fig:pitch-sweep}b.
In addition, higher order shear modes at 6~GHz are efficiently transduced.

The second set of devices fabricated consists of five delay lines in which two IDTs are separated by a distance that ranges from \(L = 100~\um\) to \(1200~\um\). The IDTs are identical with ten finger-pairs spaced by a \(1.72~\um\) pitch.
The two-port impulse response \(h\prens{\tau}\) of these delay lines is computed by discrete Fourier transform of the \(S_{21}(\omega)\), collected in Figure~\ref{fig:L-sweep}, and used to infer the propagation losses.
For devices in which the round-trip time exceeds the inverse of the bandwidth of the transducer, distinct echoes \(h_\delta\prens{\tau}\) can be identified and used to infer the round-trip loss.
The responses shown in Fig.~\ref{fig:L-sweep}a are vertically offset such that constant-velocity features align diagonally.
The SH waves propagate with group velocity \(v_\trm{g} = 4.00\pm 0.03~\trm{km}/\trm{s}\).
The intersection of the color horiztonal rules with the gray diagonals indicate 1, 2, \ldots~round-trips of the \(4000~\trm{m/s}\) SH echoes. 

If we ignore the effects of dispersion and multimodedness on \(h_\delta\prens{\tau}\) and restrict our attention to the SH band, the impulse response takes the form
\begin{equation}
	h\prens{\tau} = t^2 e^{-\alpha L/2 - i\omega \tau}\sum_{n} \prens{r^2 e^{-\alpha L}}^{n} h_\delta\prens{\tau - \tau_{n}}
	\label{eq:S21-time-domain}
\end{equation}
where the arrival time \(\tau_{n}\) is \(\prens{1 + 2n}L/v_\trm{g}\) with group velocity \(v_\trm{g}\) and \(\omega\) is the center frequency of the transducer.
Since the delay line and transducer are multimoded, the squares of the transmission coefficient \(|t|^2\) of the IDT (delay line to transmission line) and the reflection coefficient \(|r|^2\) (delay line off the IDT) do not in general add to \(1\) even in the absence of loss.
Impedance mismatch decreases \(|t|\) and increases \(|r|\).

In order to independently infer the propagation loss \(e^{-\alpha L}\) and reflection loss \(|r|^2\) from measurements, we need to sweep \(L\).
A fit to \(|h\prens{\tau_n}|\) for the \(L = 400~\um\) line alone (inset to Figure~\ref{fig:L-sweep}a) yields a round-trip loss \(|r|^2e^{-\alpha L}\) of 11.8~dB for the SH waves at \(2~\trm{GHz}\).
We independently regress \(|t|^2\), \(|r|^2\), and \(\alpha\) fitting a plane to \(|h\prens{\tau_n}|\) against \(n\) and \(nL\) for delay lines of varied length.
In doing so we assume \(t\) and \(r\) are constant across devices.
We find that the power in the \(2~\trm{GHz}\) SH0 modes drops by \(\alpha = 6.0\)~dB/mm.
This corresponds to a standing-wave resonator instrinsic quality factor of $2300$ and an $fQ$ product of $4.6\times10^{12}$.  %
This \(\alpha\) and the \(|r|^2\) loss of $7.0$~dB is consistent with the round-trip loss of $11.8$~dB inset in Figure~\ref{fig:L-sweep}a for which the round-trip length \(2L = 800~\um\).

The \(14~\um \times 1.2~\trm{mm}\) delay line has a group delay of $300$~ns with an insertion loss of $14.3$~dB, $3.4$~dB of which comes from impedance mismatch. 

The third set of devices fabricated contains 11 delay lines in which \(N\) ranges from 4 to 35.
The conductances presented in Figure~\ref{fig:N-sweep} show how the lineshape varies with \(N\) and Figure~\ref{fig:fit-results} shows the dependence of the main~lobe on \(N\).
These results reveal the influence of dispersion, reflections, and loss on the IDTs, necessitating a departure from impulse response model as discussed in Section~\ref{sec:modeling-the-conductance}.

\section{Deviation from the Impulse Response Model}
\label{sec:modeling-the-conductance}

Before accounting for loading by the electrodes, we give a brief overview of the impulse response method of IDT design~\cite{Hartmann1973}.
The linear response of the transducer is completely characterized by the static capacitance \(C_\trm{s}\) and the conductance \(G\prens{\omega}\).
The conductance \(G\prens{\omega}\) is the power spectrum of an IDT's impulse response (see Appendix~\ref{app:linear-response}) which can be related to the IDT geometry by Fourier transform.
This forms the basis of the impulse response method of transducer design.

A uniform pitch, uniformly weighted transducer driven by \(V = \delta\prens{t}\) emits an \(Na\) long rectangular pulse with wavelength \(a\) from which, assuming linear dispersion, it follows~\cite{Hartmann1973}
\begin{equation}
	G({\omega}) = G_0 \trm{sinc}^2x.
	\label{eq:sinc}
\end{equation}
Here \(x = N\pi\prens{\omega - \omega_0}/\omega_0\) and \(\omega_0 = 2\pi v/a\) where \(N\) is the number of finger-pairs, \(a\) the pitch, and \(v\) is the phase velocity of the mechanical waves.
It follows that the full-width-half-maximum \(\gamma_\trm{IRM}\) is 
\[\gamma_\trm{IRM} \approx 0.89\frac{\omega_0}{N}.\]
The peak conductance is
\[ G_0 = 8 f_0 k_\trm{eff}^2 c_\trm{s} N^2 \]
where \(k_\trm{eff}^2\) is the piezoelectric coupling coefficient and \(c_\trm{s} = C_\trm{s}/N\) is the capacitance per finger-pair.

The electrode lattice that comprises an IDT electrically and mechanically loads the piezoelectric slab.  This changes the dispersion of an IDT and introduces reflections between it and the adjoined slab.
In typical surface acoustic wave devices \(k_\trm{eff}^2\) is small and the electrodes are thin relative to the wavelength.  These conditions of low electrical and mechanical loading result in  reflections on the order of 1\% which can be treated as a second-order effect~\cite{Jones1972,morgan2010surface}. 

The devices studied here are outside the range of validity of the impulse model response model, leading to predictions that deviate significantly from measurements and observations. For example, we see that the measured and simulated bandwidth are more than an order of magnitude smaller than would be predicted by the impulse response model (see Figure~\ref{fig:fit-results}). Moreover, the scaling of the bandwidth with $N$ follows power laws different than $N^{-1}$ (see Appendix B). To understand these effects, we move to a picture that takes into account the presence of a phononic bandgap and localized resonances.

\section{The role of resonance and bandgaps}

An IDT's electrodes form a lattice that electrically and mechanically loads the slab.  This periodic loading couples forward and backward waves with wavelengths close to the pitch of the IDT, giving rise to a bandgap. In principle, a bandgap arises from an arbitrarily small periodic perturbation and can be resolved in an infinitely long, lossless IDT. Finite size and loss obscure the effects of a bandgap.
A device with loss rate $\gamma_\trm{i}$ will temporally resolve a bandgap of size $B$ when $\gamma_\trm{i}<B$. A finite-size transducer will spatially resolve the bandgap when $B$ exceeds the inverse transit time which is approximately $\gamma_\trm{IRM}$. If the bandgap is resolved both spatially and temporally, we will need to account for its effect to understand the operation of the transducer. We call this regime, in which coupling between forward and backward modes in the IDT cannot be ignored, the \textit{strong coupling} regime.

In order to capture these effects, we model a quasi-2D cross-section of our IDTs using the FEM and scale the results by the finger-overlap, $10~\um$, to compare them to measurement. 
By reducing the problem in this manner we speed up simulations while accounting for both in- and out-of-plane displacement but ignore the transverse mode structure of the IDT and slab. 
Other methods like the coupling-of-modes (COM) method and Green's function analyses also account for reflections from loading~\cite{morgan2010surface}.

\begin{figure}
    \centering
    \includegraphics[width=\linewidth]{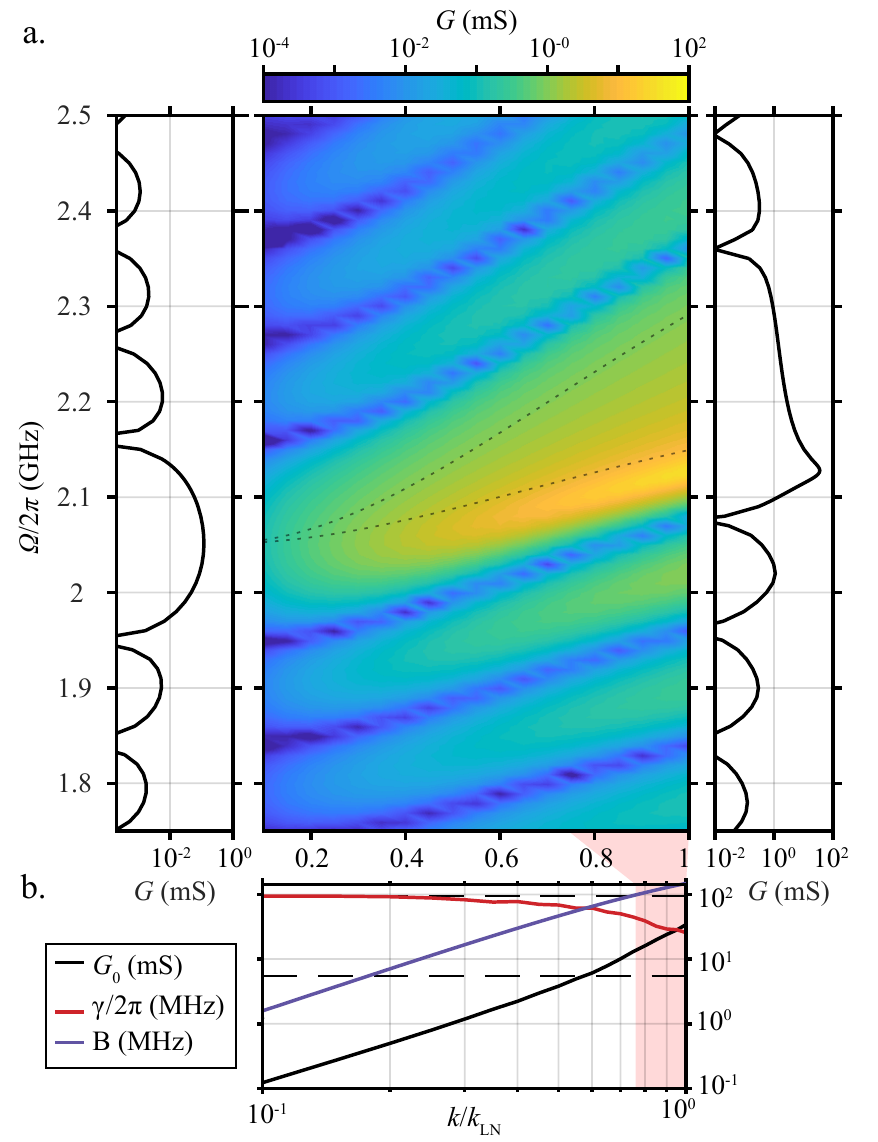}
    \caption{We scale the coefficients of the piezoelectric tensor by \(k/k_\trm{LN}\) for 10 finger-pair, double-sided IDTs with infinitely thin electrodes to isolate the effect electrical loading has on the lineshape of \(G\prens{\omega}\).  The conductance \(G\prens{\omega}\) limits smoothly between the \(\trm{sinc}^2x\) impulse response regime and the high \(k_\trm{eff}^2\) resonant regime.  The bandgap is overlaid, coinciding with the lobe-free pedestal that forms with high \(k\).  The linewidth and peak conductance of the main lobe are plotted in \textbf{b.}. The linewidth $\gamma$ is bounded above by the impulse response model and below by intrinsic losses (both dashed).  Strong coupling occurs when the splitting $B$ exceeds $\gamma_\trm{IRM}$ which happens for $N = 10$ at $k/k_\text{LN}>0.74$.}
    \label{fig:lineshape-ksquared-sweep}
\end{figure}

\begin{figure}[h]
    \centering
        \includegraphics[height=2.5in]{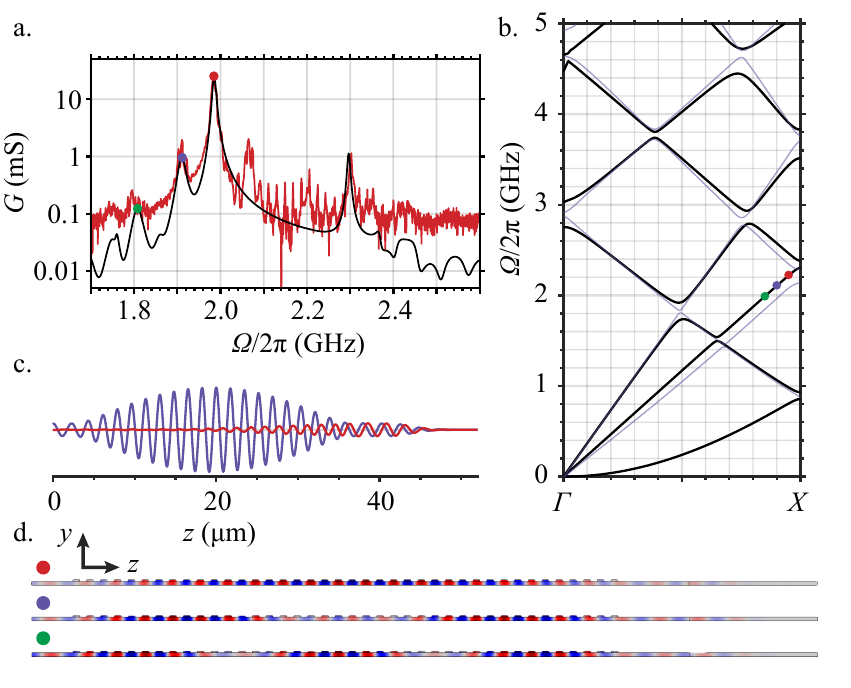}
     \caption{\textbf{a.} Measurements (red) of \(G\prens{\omega}\) are compared to lossless FEM models (black) with loss \(Q_\trm{i} = 300\) and \(k/k_\trm{LN} = 0.67\).  The lobes marked with a red, blue, and green point correspond to those on the SH0 band of a unit cell of the IDT in \textbf{b.} and the FEM solutions in \textbf{d.}. In \textbf{d.} we see the lobes can be identified by the number of nodes in the envelope.  The quadratures of the displacement \(u_x\) of the main~lobe (red point) along the bottom surface is plotted in \textbf{c.}.   The $2.3$~GHz feature in \textbf{a.} originates from a resonance near the back edge.}
     \label{fig:bands}
\end{figure}

\begin{figure}[h]
    \centering
    \includegraphics{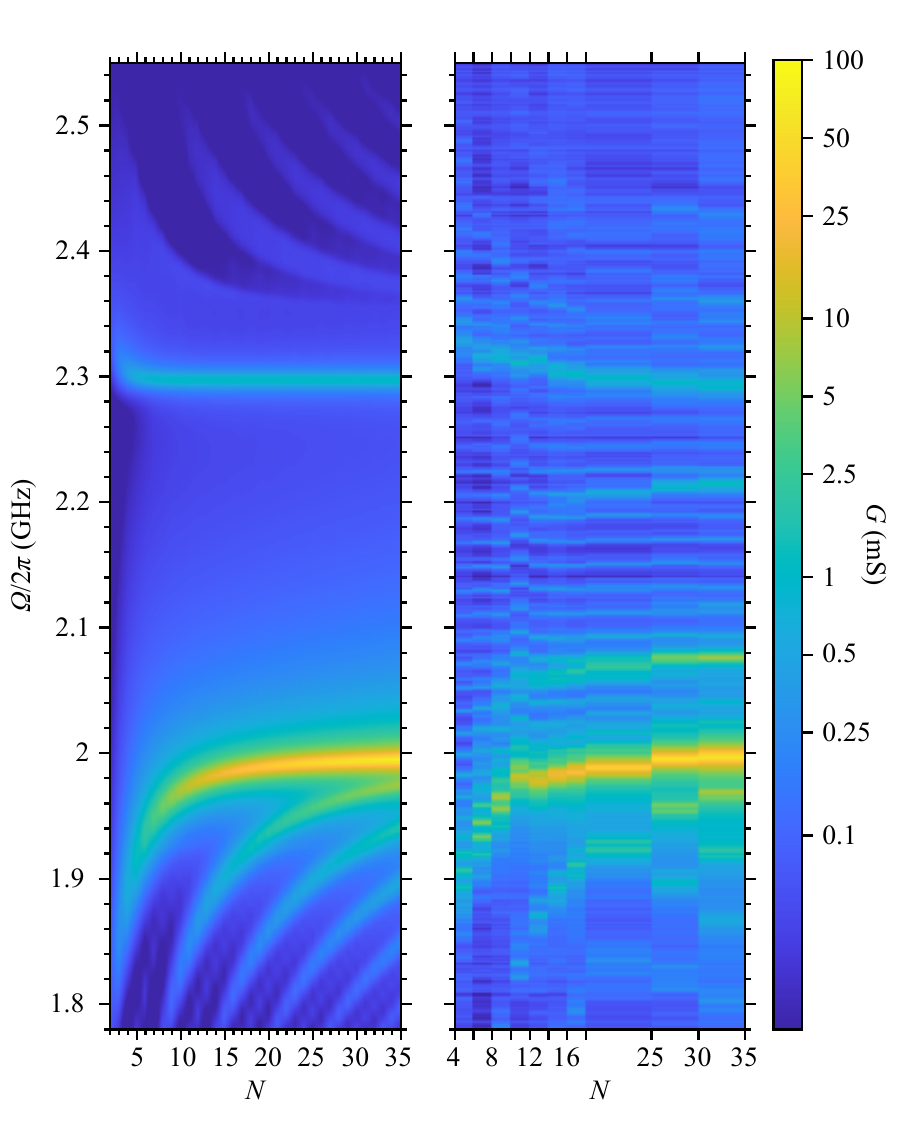}
    \caption{The simulated and measured conductance \(G\prens{\omega}\) is plotted against the number of finger-pairs \(N\) for the SH mode.  In these plots we can identify the bandgap beset below by the strongly transduced negative mass band resonances which shift up with \(N\) and above by the weakly transduced positive mass band resonances which shift down with \(N\).  Resonances along the etched back edge of the IDT appear as often strongly transduced peaks that are nearly independent of $N$. }
    \label{fig:N-sweep}
\end{figure}

First we show that for high \(k_\trm{eff}^2\) even when the electrodes are infinitely thin, electrical loading alone causes \(G\prens{\omega}\) to deviate significantly from Equation~\ref{eq:sinc}.
In Figure~\ref{fig:lineshape-ksquared-sweep} we scale the piezoelectric tensor \(d_{ij}\) by \(k/k_\trm{LN}\) between 0 and 1 and calculate \(G\prens{\omega}\) for 10 finger-pair, double-sided IDTs with infinitely thin electrodes.
As \(k_\trm{eff}^2\) approaches 0, \(G\prens{\omega}\) limits to a \(\trm{sinc}^2x\) distribution as in Equation~\ref{eq:sinc}. 
As \(k\) approaches \(k_\trm{LN}\), the lobes move up in frequency due to piezoelectric stiffening and the high frequency sidelobes shift away from the main lobe to form a large, flat pedestal.
In measurements in Figure~\ref{fig:bands}a and Figure~\ref{fig:N-sweep}b, standing waves in the delay line form Fabry-P\'erot peaks in the pedestal with a free spectral range of 20~MHz. %

In Figure~\ref{fig:lineshape-ksquared-sweep}b we plot the the simulated linewidth (full-width-half-maximum $\gamma$, solid black)  along with linewidth of the impulse response model($\gamma_\trm{IRM}$, dashed), derived from the transit time, and the intrinsic damping rate ($\gamma_\trm{i}$, dashed) which bound \(\gamma\) above and below, respectively. The bandgap $B$ increases with \(k/k_\trm{LN}\), intersecting \(\gamma_\trm{IRM}\) at \(k = 0.74 k_\trm{LN}\). Above this point, the IDT is strongly coupled for $N=10$ finger pairs. We note that above $k = 0.57 k_\trm{LN}$ the bandgap is well-resolved, exceeding the transducer linewidth $\gamma$.

The asymmetry of the lineshape in Figure~\ref{fig:lineshape-ksquared-sweep} can be understood in terms of the band structure of the electrode lattice in the high \(k_\trm{eff}^2\) limit.
In Figure~\ref{fig:bands}b, we plot the bands for infinitely thin (blue) and $100$~nm thick electrodes (black).
The avoided crossing at the X-point \(K = 2\pi/a\) suppresses the response \(G\prens{\omega}\) between the band edges forming the pedestal of Figure~\ref{fig:lineshape-ksquared-sweep}.
For ease of comparison we overlay the X-point frequencies over the domain of \(k/k_\trm{LN}\) in Figure~\ref{fig:lineshape-ksquared-sweep} (dashed).
For the IDTs measured, the $100$~nm thick aluminum electrodes also have a significant elastic effect, shifting up and reducing the size of the bandgap.

For a fixed frequency, waves propagating in the electrode lattice have a different \(K\), group velocity \(v_\trm{g}\), and mode shape than those of the free slab.
Reflections at the interface from this discontinuity result in standing waves within the IDT as shown in Figure~\ref{fig:bands}c.
In Figure~\ref{fig:bands}d we plot the fundamental, first-excited, and second-excited resonances of the negative mass band (normal group velocity dispersion \(\partial_\omega v_\trm{g} < 0\)) which correspond to the main lobe and first two low frequency sidelobes.
Likewise resonances originating from the positive mass band correspond to the high frequency sidelobes in Figures~\ref{fig:lineshape-ksquared-sweep} and \ref{fig:N-sweep}.
In varying \(k/k_\trm{LN}\) between 0 and 1 we see a smooth transition from the lobes of the impulse response which come from the \(K\) distribution of the envelope to the resonant structure of the IDT.

The strain \(S_{xz}\) of the negative and positive mass SH modes is localized in the gaps and under the electrodes, respectively, as seen in Figure~\ref{fig:pitch-sweep}b. 
These modes sample different components of the applied electric field, \(E_z\) and \(E_y\), and therefore different components of the piezoelectric tensor, \(d_\trm{YYZ}\) (\(d_{24} = d_{15}\) in Voigt notation; here capitals denote crystal axes) and \(d_\trm{XYZ} \equiv d_{14} = 0~\trm{pN}/\trm{C}\), respectively.
This contributes to the asymmetry of \(G\prens{\omega}\) and accounts for the relatively weak positive mass response in the simulated and measured conductance in Figure~\ref{fig:N-sweep}.
Changing \(d_{ij}\) by, for example, rotating the crystal changes the relative strength of these responses.

The momentum of the fundamental resonance is broadened by the spread in envelope momentum \(K_\trm{e} = 2\pi/Na\).
As \(N\) increases and \(K_\trm{e}\) approaches 0, the center frequency of the main lobe shifts toward the band edge, increasing for the negative mass modes and decreasing for the positive mass modes.
The sidelobes are similarly offset from the band edge by \(K_\trm{e}\) resulting in the $N$-dependent lobe shifts in Figure~\ref{fig:bands}a and Figure~\ref{fig:N-sweep}.

Loading affects not only the lineshape but also the shape of the lobes of \(G\prens{\omega}\).
Dispersion and reflections narrow \(\gamma\) and increase \(G_0\) of the main lobe relative to Equation~\ref{eq:sinc}.
Consider the SH response of a 16 finger-pair transducer. A \(10~\um\) wide IDT has a capacitance of 45~fF computed by the FEM.  Assuming \(k_\trm{eff}^2 = 13\%\), a \(\gamma\) of $2\pi\times110$~MHz and \(G_0\) of $1.5$~mS follows from Equation~\ref{eq:sinc}.
Instead we measure $2\pi\times 9.6$~MHz and $25$~mS, underscoring the necessity of accounting for loading.

\begin{figure}
    \centering
    \includegraphics[width=\linewidth]{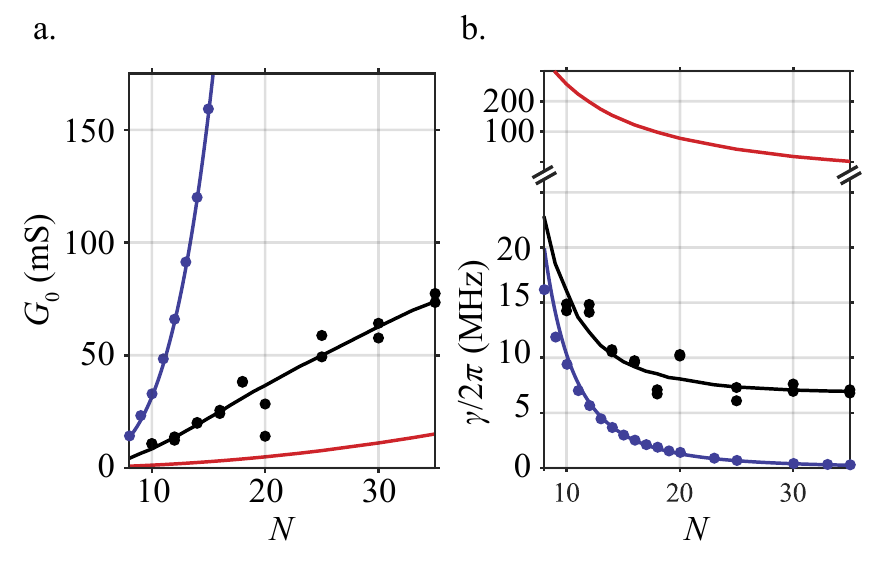}
    \caption{Lorentzians are fit to the measured and computed conductance \(G\prens{\omega}\) of Figure~\ref{fig:N-sweep} and their peak height \(G_0\) and FWHM \(\gamma\) are plotted in \textbf{a} and \textbf{b}.  Simulations without loss are plotted in blue.  By adding intrinsic damping \(\gamma_\trm{i}\) and reducing $d_{ij}$ by a factor of $0.67$, the response becomes that of the black curves.  Measured responses are plotted as black points. The red curves are predictions from the impulse response model.}
    \label{fig:fit-results}
\end{figure}

In Figure~\ref{fig:fit-results} we plot \(G_0\) and \(\gamma\) extracted from Figure~\ref{fig:N-sweep} by fitting Lorentzians to \(G\prens{\omega}\) for simulations (blue points) and measurements (black points) and compare them to Equation~\ref{eq:sinc} (red).
We simulate the system, accounting for electrical and mechanical loading, but neglecting losses. We observe an approximate \(N^{-3}\) scaling of the bandwidth $\gamma$, as opposed to the \(N^{-1}\) found in Equation~\ref{eq:sinc} and equivalent circuit models~\cite{Smith1969AnalysisModel}. As shown in the Appendix, we vary the level of electrical and mechanical loading to find a family of scaling laws between $N^{-1}$ and $N^{-3}$. In the highly coupled $N^{-3}$ limit, the stronger suppression of the bandwidth with increasing $N$ can be understood using a resonant Fabry-P\'erot model of the IDT. Waves traveling within the IDT are reflected from the IDT/slab interface with a reflection coefficient $r$ approximately determined by a group velocity mismatch \[ r = \frac{v - v_\trm{g}}{v + v_\trm{g}}, \] where $v_\trm{g}$ and $v$ are the group velocities in the IDT and slab, respectively. These reflections lead to a standing wave inside the IDT with a linewidth \[ \gamma  = \frac{2v_\trm{g}}{Na} \frac{1-r^2}{2r}\underset{r\rightarrow 1}{\rightarrow} \frac{2v_g}{Na}(1-r)= \frac{4v_g^2}{Nav}.\]
We assume that the fundamental resonance of the IDT is sufficiently close to the X-point for the dispersion to be nearly quadratic, \textit{i.e.}, \(v_\trm{g} = \beta K_\trm{e}\) for some constant \(\beta\). The spread in the envelope momentum $K_\trm{e}=2\pi/Na$ for the fundamental standing wave is inversely proportional to $N$. This leads to an expression for the  linewidth  \[ \gamma = \frac{16\pi^2\beta^2}{v a^3}\frac{1}{N^3}, \] matching the simulated power law. In addition to the original factor of \(N^{-1}\) in Equation~\ref{eq:sinc} from the length of the transducer, the group velocity dispersion and the variation in the IDT-slab reflection coefficient each contribute another factor of \(N^{-1}\). The conductance \(G_0\) is seen to scale approximately as  \(N^4\) in the same lossless simulations.

It is clear from Figure~\ref{fig:fit-results}a that the measured bandwidth \(\gamma\) stops narrowing at large $N$ and approaches $7$~MHz in the high \(N\) limit.
This intrinsic linewidth \(\gamma_\trm{i}\) corresponds to a \(Q\) of approximately $300$, significantly lower than the $Q$ of $2300$ corresponding to the $6.0$~dB/mm propagation losses measured for the $4000$~m/s SH waves.
Incorporating this loss rate into our FEM models by adding a uniform material loss tangent across the domain and reducing \(d_{ij}\) by \(k/k_\trm{LN} = 0.67\) yields the black curves in Figure~\ref{fig:fit-results}a and b in close agreement with measurements. The same material loss and reduction in \(k\) are also used in the simulations of Figure~\ref{fig:N-sweep}. 
We attribute these discrepancies to mechanical losses in the aluminum electrodes and discrepancies between the material properties in thin-film and bulk \LN, and are investigating techniques for improving the films' properties.

We note here that \(G_0\) increases at the expense of \(\gamma\) such that \(G_0\gamma\) remains proportional to \(N\).  This is a more general statement that holds regardless of the exact dependence of $\gamma$ and $G_0$ on $N$, and is a direct consequence of the extensivity of the net conductance \(\int\trm{d}\omega G\prens{\omega}\). As we will report in a future manuscript, the net conductance of a narrowband transducer can be directly related to \(k_\trm{eff}^2\) 
\[ k_\trm{eff}^2 = \frac{\int\trm{d}\omega G\prens{\omega}}{16\pi C_\trm{s} f_0^2} \]
from which we find a \(k_\trm{eff}^2\) of 10\% (area in the main lobe).
By comparing the measured static capacitance \(C_\trm{s}\) to simulations, we attribute a factor of 1.3 reduction in \(k_\trm{eff}^2\) to $22$~pF of feedthrough capacitance~\cite{Gong2013}.
For \(k = k_\trm{LN}\) the simulated \(k_\trm{eff}^2\) is $30$\%.
The discrepancy between the simulated $30$\% and measured $13$\% coincides with our choice of \(k/k_\trm{LN}\) in Figures~\ref{fig:N-sweep}a and \ref{fig:fit-results}.

\section{Conclusion}

The mechanical confinement and large \(k_\trm{eff}^2\) of piezoelectric devices in suspended \LN~slabs makes the platform an attractive candidate for a host of GHz-frequency applications.
Here we demonstrated efficient transduction of SH and longitudinal waves from 1.2 to 5.4 GHz by transducers with small footprints, \(14 \times 17~\um\).
At $2$~GHz the SH waves are strongly-coupled to the IDTs with \(k_\trm{eff}^2\) reaching 13\% (after correcting for feedthrough capacitance). 
Propagation losses for the SH waves are measured, $6.0$~dB/mm, and the waves are used to make a $1.2$~mm with a $300$~ns delay.

In this high \(k_\trm{eff}^2\) regime even in the absence of mechanical loading, electrical loading modifies the dispersion in the IDT introducing a bandgap, reflections, and resonance.
We confirmed this by studying the way the lineshape and lobe shape of the conductance \(G\prens{\omega}\) vary with the number of finger pairs \(N\) of the IDT.
The lineshape exhibits a large lobe-free pedestal in the bandgap encased by positive and negative mass band resonances which shift toward the band edge as \(N\) is increased.
Furthermore the main lobe linewidth is more strongly suppressed with increasing $N$ than in the impulse response regime. This accounts for the larger-than-expected motional admittance of these small-footprint IDTs.

For narrowband applications like the transduction of nanoresonators and waveguides for quantum devices, cavity optomechanics, or RF filters, the enhancement of the peak conductance that comes with resonance is tolerable and perhaps even desirable. Small-footprint IDTs make it easier to couple to small mechanical devices, match them to $50~\Omega$, and achieve large coupling.
But for some wideband applications in optomechanics and RF signal processing -- applications with bandwidth requirements that drive the pursuit of high \(k_\trm{eff}^2\) transducers --  reflections consequent to high \(k_\trm{eff}^2\) may pose a challenging constraint on device architecture and IDT design.
In addition to demonstrating small, high \(k_\trm{eff}^2\) transducers, our work delineates the regimes in which these effects are at play.

\section{Acknowledgements}
C.J.S. and A.S.N. would like to thank John Larson, Victor Plessky, Leonhard Reindl, Songbin Gong, Gabriel Vidal \'Alvarez, and Don and Svetlana Malocha for valuable discussions. This work was funded by the National Science Foundation through ECCS-1808100 and by the U. S. Air Force Office of Scientific Research through a MURI grant (Grant No. FA9550-17-1-0002). Part of this work was performed at the Stanford Nano Shared Facilities (SNSF) and Stanford Nanofabrication Facility (SNF) which are supported by the National Science Foundation under award ECCS-1542152. A.S.N. acknowledges the support of a David and Lucile Packard Fellowship. A.Y.C. is supported by the Stanford Graduate Fellowship program.

%\bibliography{references}

\begin{thebibliography}{24}%
	\makeatletter
	\providecommand \@ifxundefined [1]{%
		\@ifx{#1\undefined}
	}%
	\providecommand \@ifnum [1]{%
		\ifnum #1\expandafter \@firstoftwo
		\else \expandafter \@secondoftwo
		\fi
	}%
	\providecommand \@ifx [1]{%
		\ifx #1\expandafter \@firstoftwo
		\else \expandafter \@secondoftwo
		\fi
	}%
	\providecommand \natexlab [1]{#1}%
	\providecommand \enquote  [1]{``#1''}%
	\providecommand \bibnamefont  [1]{#1}%
	\providecommand \bibfnamefont [1]{#1}%
	\providecommand \citenamefont [1]{#1}%
	\providecommand \href@noop [0]{\@secondoftwo}%
	\providecommand \href [0]{\begingroup \@sanitize@url \@href}%
	\providecommand \@href[1]{\@@startlink{#1}\@@href}%
	\providecommand \@@href[1]{\endgroup#1\@@endlink}%
	\providecommand \@sanitize@url [0]{\catcode `\\12\catcode `\$12\catcode
		`\&12\catcode `\#12\catcode `\^12\catcode `\_12\catcode `\%12\relax}%
	\providecommand \@@startlink[1]{}%
	\providecommand \@@endlink[0]{}%
	\providecommand \url  [0]{\begingroup\@sanitize@url \@url }%
	\providecommand \@url [1]{\endgroup\@href {#1}{\urlprefix }}%
	\providecommand \urlprefix  [0]{URL }%
	\providecommand \Eprint [0]{\href }%
	\providecommand \doibase [0]{http://dx.doi.org/}%
	\providecommand \selectlanguage [0]{\@gobble}%
	\providecommand \bibinfo  [0]{\@secondoftwo}%
	\providecommand \bibfield  [0]{\@secondoftwo}%
	\providecommand \translation [1]{[#1]}%
	\providecommand \BibitemOpen [0]{}%
	\providecommand \bibitemStop [0]{}%
	\providecommand \bibitemNoStop [0]{.\EOS\space}%
	\providecommand \EOS [0]{\spacefactor3000\relax}%
	\providecommand \BibitemShut  [1]{\csname bibitem#1\endcsname}%
	\let\auto@bib@innerbib\@empty
	%</preamble>
	\bibitem [{\citenamefont {Morgan}(2002)}]{Morgan2002}%
	\BibitemOpen
	\bibfield  {author} {\bibinfo {author} {\bibfnamefont {D.}~\bibnamefont
			{Morgan}},\ }in\ \href {\doibase 10.1109/FREQ.1998.717937} {\emph {\bibinfo
			{booktitle} {Proceedings of the 1998 IEEE International Frequency Control
				Symposium (Cat. No.98CH36165)}}}\ (\bibinfo  {publisher} {IEEE},\ \bibinfo
	{year} {2002})\ pp.\ \bibinfo {pages} {439--460}\BibitemShut {NoStop}%
	\bibitem [{\citenamefont {Olsson}\ \emph {et~al.}(2014)\citenamefont {Olsson},
		\citenamefont {Hattar}, \citenamefont {Homeijer}, \citenamefont {Wiwi},
		\citenamefont {Eichenfield}, \citenamefont {Branch}, \citenamefont {Baker},
		\citenamefont {Nguyen}, \citenamefont {Clark}, \citenamefont {Bauer},\ and\
		\citenamefont {Friedmann}}]{Olsson2014}%
	\BibitemOpen
	\bibfield  {author} {\bibinfo {author} {\bibfnamefont {R.~H.}\ \bibnamefont
			{Olsson}}, \bibinfo {author} {\bibfnamefont {K.}~\bibnamefont {Hattar}},
		\bibinfo {author} {\bibfnamefont {S.~J.}\ \bibnamefont {Homeijer}}, \bibinfo
		{author} {\bibfnamefont {M.}~\bibnamefont {Wiwi}}, \bibinfo {author}
		{\bibfnamefont {M.}~\bibnamefont {Eichenfield}}, \bibinfo {author}
		{\bibfnamefont {D.~W.}\ \bibnamefont {Branch}}, \bibinfo {author}
		{\bibfnamefont {M.~S.}\ \bibnamefont {Baker}}, \bibinfo {author}
		{\bibfnamefont {J.}~\bibnamefont {Nguyen}}, \bibinfo {author} {\bibfnamefont
			{B.}~\bibnamefont {Clark}}, \bibinfo {author} {\bibfnamefont
			{T.}~\bibnamefont {Bauer}}, \ and\ \bibinfo {author} {\bibfnamefont {T.~A.}\
			\bibnamefont {Friedmann}},\ }\href {\doibase 10.1016/j.sna.2014.01.033}
	{\bibfield  {journal} {\bibinfo  {journal} {Sensors and Actuators A:
				Physical}\ }\textbf {\bibinfo {volume} {209}},\ \bibinfo {pages} {183}
		(\bibinfo {year} {2014})}\BibitemShut {NoStop}%
	\bibitem [{\citenamefont {Song}\ \emph {et~al.}(2016)\citenamefont {Song},
		\citenamefont {Lu},\ and\ \citenamefont {Gong}}]{Song2016}%
	\BibitemOpen
	\bibfield  {author} {\bibinfo {author} {\bibfnamefont {Y.~H.}\ \bibnamefont
			{Song}}, \bibinfo {author} {\bibfnamefont {R.}~\bibnamefont {Lu}}, \ and\
		\bibinfo {author} {\bibfnamefont {S.}~\bibnamefont {Gong}},\ }\href {\doibase
		10.1109/TED.2016.2543742} {\bibfield  {journal} {\bibinfo  {journal} {IEEE
				Transactions on Electron Devices}\ }\textbf {\bibinfo {volume} {63}},\
		\bibinfo {pages} {2066} (\bibinfo {year} {2016})}\BibitemShut {NoStop}%
	\bibitem [{\citenamefont {Manzaneque}\ \emph {et~al.}(2017)\citenamefont
		{Manzaneque}, \citenamefont {Lu}, \citenamefont {Yang},\ and\ \citenamefont
		{Gong}}]{Manzaneque2017}%
	\BibitemOpen
	\bibfield  {author} {\bibinfo {author} {\bibfnamefont {T.}~\bibnamefont
			{Manzaneque}}, \bibinfo {author} {\bibfnamefont {R.}~\bibnamefont {Lu}},
		\bibinfo {author} {\bibfnamefont {Y.}~\bibnamefont {Yang}}, \ and\ \bibinfo
		{author} {\bibfnamefont {S.}~\bibnamefont {Gong}},\ }\href {\doibase
		10.1109/JMEMS.2017.2750176} {\bibfield  {journal} {\bibinfo  {journal}
			{Journal of Microelectromechanical Systems}\ }\textbf {\bibinfo {volume}
			{26}},\ \bibinfo {pages} {1204} (\bibinfo {year} {2017})}\BibitemShut
	{NoStop}%
	\bibitem [{\citenamefont {Gong}\ \emph {et~al.}(2017)\citenamefont {Gong},
		\citenamefont {Song}, \citenamefont {Manzaneque}, \citenamefont {Lu},
		\citenamefont {Yang},\ and\ \citenamefont {Kourani}}]{Gong2017}%
	\BibitemOpen
	\bibfield  {author} {\bibinfo {author} {\bibfnamefont {S.}~\bibnamefont
			{Gong}}, \bibinfo {author} {\bibfnamefont {Y.-H.}\ \bibnamefont {Song}},
		\bibinfo {author} {\bibfnamefont {T.}~\bibnamefont {Manzaneque}}, \bibinfo
		{author} {\bibfnamefont {R.}~\bibnamefont {Lu}}, \bibinfo {author}
		{\bibfnamefont {Y.}~\bibnamefont {Yang}}, \ and\ \bibinfo {author}
		{\bibfnamefont {A.}~\bibnamefont {Kourani}},\ }in\ \href {\doibase
		10.1109/MWSCAS.2017.8052856} {\emph {\bibinfo {booktitle} {2017 IEEE 60th
				International Midwest Symposium on Circuits and Systems (MWSCAS)}}}\
	(\bibinfo  {publisher} {IEEE},\ \bibinfo {year} {2017})\ pp.\ \bibinfo
	{pages} {45--48}\BibitemShut {NoStop}%
	\bibitem [{\citenamefont {Sarabalis}\ \emph {et~al.}(2018)\citenamefont
		{Sarabalis}, \citenamefont {Van~Laer},\ and\ \citenamefont
		{Safavi-Naeini}}]{Sarabalis2018}%
	\BibitemOpen
	\bibfield  {author} {\bibinfo {author} {\bibfnamefont {C.~J.}\ \bibnamefont
			{Sarabalis}}, \bibinfo {author} {\bibfnamefont {R.}~\bibnamefont {Van~Laer}},
		\ and\ \bibinfo {author} {\bibfnamefont {A.~H.}\ \bibnamefont
			{Safavi-Naeini}},\ }\href {\doibase 10.1364/OE.26.022075} {\bibfield
		{journal} {\bibinfo  {journal} {Optics Express}\ }\textbf {\bibinfo {volume}
			{26}},\ \bibinfo {pages} {22075} (\bibinfo {year} {2018})}\BibitemShut
	{NoStop}%
	\bibitem [{\citenamefont {Safavi-Naeini}\ \emph {et~al.}(2019)\citenamefont
		{Safavi-Naeini}, \citenamefont {Van~Thourhout}, \citenamefont {Baets},\ and\
		\citenamefont {Van~Laer}}]{VanLaer2019}%
	\BibitemOpen
	\bibfield  {author} {\bibinfo {author} {\bibfnamefont {A.~H.}\ \bibnamefont
			{Safavi-Naeini}}, \bibinfo {author} {\bibfnamefont {D.}~\bibnamefont
			{Van~Thourhout}}, \bibinfo {author} {\bibfnamefont {R.}~\bibnamefont
			{Baets}}, \ and\ \bibinfo {author} {\bibfnamefont {R.}~\bibnamefont
			{Van~Laer}},\ }\href {\doibase 10.1364/OPTICA.6.000213} {\bibfield  {journal}
		{\bibinfo  {journal} {Optica}\ }\textbf {\bibinfo {volume} {6}},\ \bibinfo
		{pages} {213} (\bibinfo {year} {2019})}\BibitemShut {NoStop}%
	\bibitem [{\citenamefont {Mahmoud}\ \emph {et~al.}(2018)\citenamefont
		{Mahmoud}, \citenamefont {Mahmoud}, \citenamefont {Cai}, \citenamefont
		{Khan}, \citenamefont {Mukherjee}, \citenamefont {Bain},\ and\ \citenamefont
		{Piazza}}]{Khan2018}%
	\BibitemOpen
	\bibfield  {author} {\bibinfo {author} {\bibfnamefont {M.}~\bibnamefont
			{Mahmoud}}, \bibinfo {author} {\bibfnamefont {A.}~\bibnamefont {Mahmoud}},
		\bibinfo {author} {\bibfnamefont {L.}~\bibnamefont {Cai}}, \bibinfo {author}
		{\bibfnamefont {M.}~\bibnamefont {Khan}}, \bibinfo {author} {\bibfnamefont
			{T.}~\bibnamefont {Mukherjee}}, \bibinfo {author} {\bibfnamefont
			{J.}~\bibnamefont {Bain}}, \ and\ \bibinfo {author} {\bibfnamefont
			{G.}~\bibnamefont {Piazza}},\ }\href {\doibase 10.1364/OE.26.025060}
	{\bibfield  {journal} {\bibinfo  {journal} {Optics Express}\ }\textbf
		{\bibinfo {volume} {26}},\ \bibinfo {pages} {25060} (\bibinfo {year}
		{2018})}\BibitemShut {NoStop}%
	\bibitem [{\citenamefont {Kittlaus}\ \emph {et~al.}(2016)\citenamefont
		{Kittlaus}, \citenamefont {Shin},\ and\ \citenamefont
		{Rakich}}]{Kittlaus2015}%
	\BibitemOpen
	\bibfield  {author} {\bibinfo {author} {\bibfnamefont {E.~A.}\ \bibnamefont
			{Kittlaus}}, \bibinfo {author} {\bibfnamefont {H.}~\bibnamefont {Shin}}, \
		and\ \bibinfo {author} {\bibfnamefont {P.~T.}\ \bibnamefont {Rakich}},\
	}\href {\doibase 10.1038/nphoton.2016.112} {\bibfield  {journal} {\bibinfo
			{journal} {Nature Photonics}\ }\textbf {\bibinfo {volume} {10}},\ \bibinfo
		{pages} {463} (\bibinfo {year} {2016})}\BibitemShut {NoStop}%
	\bibitem [{\citenamefont {Sohn}\ \emph {et~al.}(2018)\citenamefont {Sohn},
		\citenamefont {Kim},\ and\ \citenamefont {Bahl}}]{Sohn2018}%
	\BibitemOpen
	\bibfield  {author} {\bibinfo {author} {\bibfnamefont {D.~B.}\ \bibnamefont
			{Sohn}}, \bibinfo {author} {\bibfnamefont {S.}~\bibnamefont {Kim}}, \ and\
		\bibinfo {author} {\bibfnamefont {G.}~\bibnamefont {Bahl}},\ }in\ \href
	{\doibase 10.1109/IPCon.2018.8527332} {\emph {\bibinfo {booktitle} {2018 IEEE
				Photonics Conference (IPC)}}}\ (\bibinfo  {publisher} {IEEE},\ \bibinfo
	{year} {2018})\ pp.\ \bibinfo {pages} {1--2}\BibitemShut {NoStop}%
	\bibitem [{\citenamefont {Tadesse}\ and\ \citenamefont
		{Li}(2014)}]{Tadesse2014}%
	\BibitemOpen
	\bibfield  {author} {\bibinfo {author} {\bibfnamefont {S.~A.}\ \bibnamefont
			{Tadesse}}\ and\ \bibinfo {author} {\bibfnamefont {M.}~\bibnamefont {Li}},\
	}\href {\doibase 10.1038/ncomms6402} {\bibfield  {journal} {\bibinfo
			{journal} {Nature Communications}\ }\textbf {\bibinfo {volume} {5}},\
		\bibinfo {pages} {5402} (\bibinfo {year} {2014})}\BibitemShut {NoStop}%
	\bibitem [{\citenamefont {Satzinger}\ \emph {et~al.}(2018)\citenamefont
		{Satzinger}, \citenamefont {Zhong}, \citenamefont {Chang}, \citenamefont
		{Peairs}, \citenamefont {Bienfait}, \citenamefont {Chou}, \citenamefont
		{Cleland}, \citenamefont {Conner}, \citenamefont {{Dumur}}, \citenamefont
		{Grebel}, \citenamefont {Gutierrez}, \citenamefont {November}, \citenamefont
		{Povey}, \citenamefont {Whiteley}, \citenamefont {Awschalom}, \citenamefont
		{Schuster},\ and\ \citenamefont {Cleland}}]{Satzinger2018QuantumPhonons}%
	\BibitemOpen
	\bibfield  {author} {\bibinfo {author} {\bibfnamefont {K.~J.}\ \bibnamefont
			{Satzinger}}, \bibinfo {author} {\bibfnamefont {Y.~P.}\ \bibnamefont
			{Zhong}}, \bibinfo {author} {\bibfnamefont {H.~S.}\ \bibnamefont {Chang}},
		\bibinfo {author} {\bibfnamefont {G.~A.}\ \bibnamefont {Peairs}}, \bibinfo
		{author} {\bibfnamefont {A.}~\bibnamefont {Bienfait}}, \bibinfo {author}
		{\bibfnamefont {M.~H.}\ \bibnamefont {Chou}}, \bibinfo {author}
		{\bibfnamefont {A.~Y.}\ \bibnamefont {Cleland}}, \bibinfo {author}
		{\bibfnamefont {C.~R.}\ \bibnamefont {Conner}}, \bibinfo {author}
		{\bibnamefont {{Dumur}}}, \bibinfo {author} {\bibfnamefont {J.}~\bibnamefont
			{Grebel}}, \bibinfo {author} {\bibfnamefont {I.}~\bibnamefont {Gutierrez}},
		\bibinfo {author} {\bibfnamefont {B.~H.}\ \bibnamefont {November}}, \bibinfo
		{author} {\bibfnamefont {R.~G.}\ \bibnamefont {Povey}}, \bibinfo {author}
		{\bibfnamefont {S.~J.}\ \bibnamefont {Whiteley}}, \bibinfo {author}
		{\bibfnamefont {D.~D.}\ \bibnamefont {Awschalom}}, \bibinfo {author}
		{\bibfnamefont {D.~I.}\ \bibnamefont {Schuster}}, \ and\ \bibinfo {author}
		{\bibfnamefont {A.~N.}\ \bibnamefont {Cleland}},\ }\href {\doibase
		10.1038/s41586-018-0719-5} {\bibfield  {journal} {\bibinfo  {journal}
			{Nature}\ }\textbf {\bibinfo {volume} {563}},\ \bibinfo {pages} {661}
		(\bibinfo {year} {2018})}\BibitemShut {NoStop}%
	\bibitem [{\citenamefont {Pechal}\ \emph {et~al.}(2018)\citenamefont {Pechal},
		\citenamefont {Arrangoiz-Arriola},\ and\ \citenamefont
		{Safavi-Naeini}}]{Pechal2018SuperconductingStorage}%
	\BibitemOpen
	\bibfield  {author} {\bibinfo {author} {\bibfnamefont {M.}~\bibnamefont
			{Pechal}}, \bibinfo {author} {\bibfnamefont {P.}~\bibnamefont
			{Arrangoiz-Arriola}}, \ and\ \bibinfo {author} {\bibfnamefont {A.~H.}\
			\bibnamefont {Safavi-Naeini}},\ }\href {\doibase 10.1088/2058-9565/aadc6c}
	{\bibfield  {journal} {\bibinfo  {journal} {Quantum Science and Technology}\
		}\textbf {\bibinfo {volume} {4}},\ \bibinfo {pages} {015006} (\bibinfo {year}
		{2018})}\BibitemShut {NoStop}%
	\bibitem [{\citenamefont {Kuznetsova}\ \emph {et~al.}(2001)\citenamefont
		{Kuznetsova}, \citenamefont {Zaitsev}, \citenamefont {Joshi},\ and\
		\citenamefont {Borodina}}]{Zaitsev2001}%
	\BibitemOpen
	\bibfield  {author} {\bibinfo {author} {\bibfnamefont {I.}~\bibnamefont
			{Kuznetsova}}, \bibinfo {author} {\bibfnamefont {B.}~\bibnamefont {Zaitsev}},
		\bibinfo {author} {\bibfnamefont {S.}~\bibnamefont {Joshi}}, \ and\ \bibinfo
		{author} {\bibfnamefont {I.}~\bibnamefont {Borodina}},\ }\href {\doibase
		10.1109/58.896145} {\bibfield  {journal} {\bibinfo  {journal} {IEEE
				Transactions on Ultrasonics, Ferroelectrics and Frequency Control}\ }\textbf
		{\bibinfo {volume} {48}},\ \bibinfo {pages} {322} (\bibinfo {year}
		{2001})}\BibitemShut {NoStop}%
	\bibitem [{\citenamefont {Pop}\ \emph {et~al.}(2017)\citenamefont {Pop},
		\citenamefont {Kochhar}, \citenamefont {Vidal-Alvarez},\ and\ \citenamefont
		{Piazza}}]{Pop2017}%
	\BibitemOpen
	\bibfield  {author} {\bibinfo {author} {\bibfnamefont {F.~V.}\ \bibnamefont
			{Pop}}, \bibinfo {author} {\bibfnamefont {A.~S.}\ \bibnamefont {Kochhar}},
		\bibinfo {author} {\bibfnamefont {G.}~\bibnamefont {Vidal-Alvarez}}, \ and\
		\bibinfo {author} {\bibfnamefont {G.}~\bibnamefont {Piazza}},\ }in\ \href
	{\doibase 10.1109/MEMSYS.2017.7863571} {\emph {\bibinfo {booktitle} {2017
				IEEE 30th International Conference on Micro Electro Mechanical Systems
				(MEMS)}}}\ (\bibinfo  {publisher} {IEEE},\ \bibinfo {year} {2017})\ pp.\
	\bibinfo {pages} {966--969}\BibitemShut {NoStop}%
	\bibitem [{\citenamefont {Vidal-Alvarez}\ \emph {et~al.}(2017)\citenamefont
		{Vidal-Alvarez}, \citenamefont {Kochhar},\ and\ \citenamefont
		{Piazza}}]{Vidal-alvarez2017}%
	\BibitemOpen
	\bibfield  {author} {\bibinfo {author} {\bibfnamefont {G.}~\bibnamefont
			{Vidal-Alvarez}}, \bibinfo {author} {\bibfnamefont {A.}~\bibnamefont
			{Kochhar}}, \ and\ \bibinfo {author} {\bibfnamefont {G.}~\bibnamefont
			{Piazza}},\ }in\ \href {\doibase 10.1109/ULTSYM.2017.8091845} {\emph
		{\bibinfo {booktitle} {2017 IEEE International Ultrasonics Symposium
				(IUS)}}}\ (\bibinfo  {publisher} {IEEE},\ \bibinfo {year} {2017})\ pp.\
	\bibinfo {pages} {1--4}\BibitemShut {NoStop}%
	\bibitem [{\citenamefont {Weis}\ and\ \citenamefont
		{Gaylord}(1985)}]{Weis1985}%
	\BibitemOpen
	\bibfield  {author} {\bibinfo {author} {\bibfnamefont {R.~S.}\ \bibnamefont
			{Weis}}\ and\ \bibinfo {author} {\bibfnamefont {T.~K.}\ \bibnamefont
			{Gaylord}},\ }\href {\doibase 10.1007/BF00614817} {\bibfield  {journal}
		{\bibinfo  {journal} {Applied Physics A Solids and Surfaces}\ }\textbf
		{\bibinfo {volume} {37}},\ \bibinfo {pages} {191} (\bibinfo {year}
		{1985})}\BibitemShut {NoStop}%
	\bibitem [{\citenamefont {Arrangoiz-Arriola}\ \emph {et~al.}(2018)\citenamefont
		{Arrangoiz-Arriola}, \citenamefont {Wollack}, \citenamefont {Pechal},
		\citenamefont {Witmer}, \citenamefont {Hill},\ and\ \citenamefont
		{Safavi-Naeini}}]{Arrangoiz-Arriola2018}%
	\BibitemOpen
	\bibfield  {author} {\bibinfo {author} {\bibfnamefont {P.}~\bibnamefont
			{Arrangoiz-Arriola}}, \bibinfo {author} {\bibfnamefont {E.~A.}\ \bibnamefont
			{Wollack}}, \bibinfo {author} {\bibfnamefont {M.}~\bibnamefont {Pechal}},
		\bibinfo {author} {\bibfnamefont {J.~D.}\ \bibnamefont {Witmer}}, \bibinfo
		{author} {\bibfnamefont {J.~T.}\ \bibnamefont {Hill}}, \ and\ \bibinfo
		{author} {\bibfnamefont {A.~H.}\ \bibnamefont {Safavi-Naeini}},\ }\href
	{\doibase 10.1103/PhysRevX.8.031007} {\bibfield  {journal} {\bibinfo
			{journal} {Physical Review X}\ }\textbf {\bibinfo {volume} {8}},\ \bibinfo
		{pages} {031007} (\bibinfo {year} {2018})}\BibitemShut {NoStop}%
	\bibitem [{\citenamefont {Pozar}(2009)}]{Pozar2009}%
	\BibitemOpen
	\bibfield  {author} {\bibinfo {author} {\bibfnamefont {D.~M.}\ \bibnamefont
			{Pozar}},\ }\href@noop {} {\emph {\bibinfo {title} {{Microwave
					engineering}}}},\ \bibinfo {edition} {3rd}\ ed.\ (\bibinfo  {publisher} {John
		Wiley {\&} Sons, Inc.},\ \bibinfo {year} {2009})\BibitemShut {NoStop}%
	\bibitem [{\citenamefont {Hartmann}\ \emph {et~al.}(1973)\citenamefont
		{Hartmann}, \citenamefont {Bell},\ and\ \citenamefont
		{Rosenfeld}}]{Hartmann1973}%
	\BibitemOpen
	\bibfield  {author} {\bibinfo {author} {\bibfnamefont {C.}~\bibnamefont
			{Hartmann}}, \bibinfo {author} {\bibfnamefont {D.}~\bibnamefont {Bell}}, \
		and\ \bibinfo {author} {\bibfnamefont {R.}~\bibnamefont {Rosenfeld}},\ }\href
	{\doibase 10.1109/TMTT.1973.1127967} {\bibfield  {journal} {\bibinfo
			{journal} {IEEE Transactions on Microwave Theory and Techniques}\ }\textbf
		{\bibinfo {volume} {21}},\ \bibinfo {pages} {162} (\bibinfo {year}
		{1973})}\BibitemShut {NoStop}%
	\bibitem [{\citenamefont {Jones}\ \emph {et~al.}(1972)\citenamefont {Jones},
		\citenamefont {Hartman},\ and\ \citenamefont {Sturdivant}}]{Jones1972}%
	\BibitemOpen
	\bibfield  {author} {\bibinfo {author} {\bibfnamefont {W.}~\bibnamefont
			{Jones}}, \bibinfo {author} {\bibfnamefont {C.}~\bibnamefont {Hartman}}, \
		and\ \bibinfo {author} {\bibfnamefont {T.}~\bibnamefont {Sturdivant}},\
	}\href {\doibase 10.1109/T-SU.1972.29688} {\bibfield  {journal} {\bibinfo
			{journal} {IEEE Transactions on Sonics and Ultrasonics}\ }\textbf {\bibinfo
			{volume} {19}},\ \bibinfo {pages} {368} (\bibinfo {year} {1972})}\BibitemShut
	{NoStop}%
	\bibitem [{\citenamefont {Morgan}(2010)}]{morgan2010surface}%
	\BibitemOpen
	\bibfield  {author} {\bibinfo {author} {\bibfnamefont {D.~P.}\ \bibnamefont
			{Morgan}},\ }\href@noop {} {\emph {\bibinfo {title} {{Surface acoustic wave
					filters: With applications to electronic communications and signal
					processing}}}}\ (\bibinfo  {publisher} {Academic Press},\ \bibinfo {year}
	{2010})\BibitemShut {NoStop}%
	\bibitem [{\citenamefont {Smith}\ \emph {et~al.}(1969)\citenamefont {Smith},
		\citenamefont {Gerard}, \citenamefont {Collins}, \citenamefont {Reeder},\
		and\ \citenamefont {Shaw}}]{Smith1969AnalysisModel}%
	\BibitemOpen
	\bibfield  {author} {\bibinfo {author} {\bibfnamefont {W.}~\bibnamefont
			{Smith}}, \bibinfo {author} {\bibfnamefont {H.}~\bibnamefont {Gerard}},
		\bibinfo {author} {\bibfnamefont {J.}~\bibnamefont {Collins}}, \bibinfo
		{author} {\bibfnamefont {T.}~\bibnamefont {Reeder}}, \ and\ \bibinfo {author}
		{\bibfnamefont {H.}~\bibnamefont {Shaw}},\ }\href {\doibase
		10.1109/TMTT.1969.1127075} {\bibfield  {journal} {\bibinfo  {journal} {IEEE
				Transactions on Microwave Theory and Techniques}\ }\textbf {\bibinfo {volume}
			{17}},\ \bibinfo {pages} {856} (\bibinfo {year} {1969})}\BibitemShut
	{NoStop}%
	\bibitem [{\citenamefont {Gong}\ and\ \citenamefont {Piazza}(2013)}]{Gong2013}%
	\BibitemOpen
	\bibfield  {author} {\bibinfo {author} {\bibfnamefont {S.}~\bibnamefont
			{Gong}}\ and\ \bibinfo {author} {\bibfnamefont {G.}~\bibnamefont {Piazza}},\
	}\href {\doibase 10.1109/TMTT.2012.2228671} {\bibfield  {journal} {\bibinfo
			{journal} {IEEE Transactions on Microwave Theory and Techniques}\ }\textbf
		{\bibinfo {volume} {61}},\ \bibinfo {pages} {403} (\bibinfo {year}
		{2013})}\BibitemShut {NoStop}%
\end{thebibliography}

\appendix 
\section{Linear response of a transducer}
\label{app:linear-response}
Consider an \(N\) finger-pair transducer with anode-to-anode spacing, or \emph{pitch}, \(a\) and uniform finger overlap, or \emph{weighting}, \(w\).
The transducer is a linear time-invariant system and is thus described within the context of linear response theory.
Applying a voltage \(V\prens{t^\prime}\) across the device draws a current \(I\prens{t}\) related to \(V\) by the impulse response function \(Y\prens{t - t^\prime}\) 
\[ I\prens{t} = \int_{-\infty}^\infty\trm{d}t^\prime Y\prens{t - t^\prime} V\prens{t^\prime}. \]
A \(\delta\)-distributed drive \(V\prens{t^\prime} = \delta\prens{t^\prime}\) generates a current \(Y\prens{t}\) . 

For time-invariant systems where the impulse response depends only on the difference \(t - t^\prime\), it takes a simple form in the frequency domain
\begin{equation}
    I\prens{\omega} = Y\prens{\omega} V\prens{\omega}
    \label{eq:admittance}
\end{equation}
where 
\[ V\prens{t} = \frac{1}{2\pi}\int_{-\infty}^{\infty}\trm{d}\omega V\prens{\omega}e^{-i\omega t} \]
and similarly for the current \(I\).
The impulse response \(Y\prens{t}\) and circuit admittance \(Y\prens{\omega}\) are related by a Fourier transform.
Since the voltage \(V\prens{t}\) is a real-valued function, the complex amplitude \(V\prens{-\omega}\) is equal to \(V^*\prens{\omega}\), similarly for \(I\prens{\omega}\), and therefore \(Y\prens{\omega}\).

For a single port device such as an IDT terminating a transmission line, measurements of reflections encoded in the scattering parameter \(S_{11}\prens{\omega}\) can be used to compute the admittance
\[ Y\prens{\omega} = Y_0 \frac{1 - S_{11}\prens{\omega}}{1 + S_{11}\prens{\omega}} \]
where \(Y_0\) is the admittance of the transmission line, often 20~mS.
For a two-port device, the linear response can be captured by an admittance matrix with components \(Y_{ij}\) which can be computed from elements of the scattering matrix \(S_{ij}\)~\cite{Pozar2009}.
The scattering matrix can be measured using a vector network analyzer from which one can compute the impulse response
\[ Y\prens{t} = \frac{1}{2\pi}\int_{-\infty}^\infty\trm{d}\omega Y\prens{\omega} e^{-i\omega t} \]
with units of \(\trm{Hz}/\Omega\).
A lot of the information about the response of a system is encoded in the phase of \(S_{11}\) which often has a simpler structure and interpretation in the time-domain as shown in Section~\ref{sec:S-band-IDTs}.

The real part of the admittance, the \emph{conductance} \(G\prens{\omega}\), is related directly to the time-averaged power dissipated by a system.
The power dissipated by a system \(\mathcal{P}\) is the product of the voltage and current; therefore the spectrum of the dissipated power is the convolution of \(V\prens{\omega}\) and \(I\prens{\omega}\).
Averaging in time extracts the DC component of the power \(\mathcal{P}_0\), reducing the convolution to the cross product \(V\prens{\omega} I\prens{-\omega}+ V\prens{-\omega}I\prens{\omega}\) integrated over positive \(\omega\).
Thus we find
\begin{align}
	\mathcal{P}_0 = \int_{-\infty}^\infty\trm{d}\omega G\prens{\omega} \left| V\prens{\omega}\right|^2 
		\label{eq:time-averaged-power}
\end{align}
which is a weighted average of the power spectral density of \(V\).
Causality constrains the linear response such that the conductance and the imaginary part of \(Y\prens{\omega}\), the \emph{susceptance}, are a Hilbert transform pair and therefore not independent quantities.  Computing \(G\prens{\omega}\) and the electrostatic capacitance \(C_\trm{s}\) is a complete description of the linear response.

If we ignore material losses, resistive losses, and microwave radiation, the power dissipated by the IDT radiates away as mechanical waves in the lithium niobate.
Because we can relate the power-dissipated to the radiated power, we can directly compute the conductance \(G\prens{\omega}\) from the power-spectrum of the mechanical fields impulse-response. This forms the basis of the impulse response method~\cite{Hartmann1973}.

\section{Dependence of the linewidth and peak conductance on mechanical and electrical loading}
\label{app:electrode-thickness}

It's reasonable to suspect that the resonance and reflections arise predominately from the mechanical, not electrical, loading by the electrodes.  The electrodes are 100~nm thick, a significant fraction of the 250~nm \LN~slab.
If this was the case, as the electrode thickness approaches 0, the peak conductance \(G_0\) would scale as \(N^2\) and the FWHM \(\gamma\) as $N^{-1}$ as in Equation~\ref{eq:sinc} of the impulse response model.

To investigate this limit, we compute the conductance \(G\prens{\omega}\) for IDTs between 10 and 18 finger-pairs, scaling the piezoelectric tensor \(\mathbf{d}\) by \(k/k_\trm{LN}\) between $0.1$ to $1$ and the electrode thickness from 0 to 100~nm.
Lorentzians are fit to \(G\prens{\omega}\) and the resulting parameters are fit against \(N\) on a log-log plot.
The exponents $x$ are plotted in the form
\begin{align}
    G_0 &\propto N^{2+x} \\
    \gamma &\propto N^{-1-x}.
\end{align}
The \(N^{-3}\) high \(t_\trm{Al}\) and \(k_\trm{eff}^2\) limit of \(\gamma\) is given context in Section~\ref{sec:modeling-the-conductance}.
Even when \(t_\trm{Al} = 0~\trm{nm}\), \(\gamma\) and \(G_0\) deviate from the power-law scaling of Equation~\ref{eq:sinc}.

\begin{figure}
    \centering
    \includegraphics[width=\linewidth]{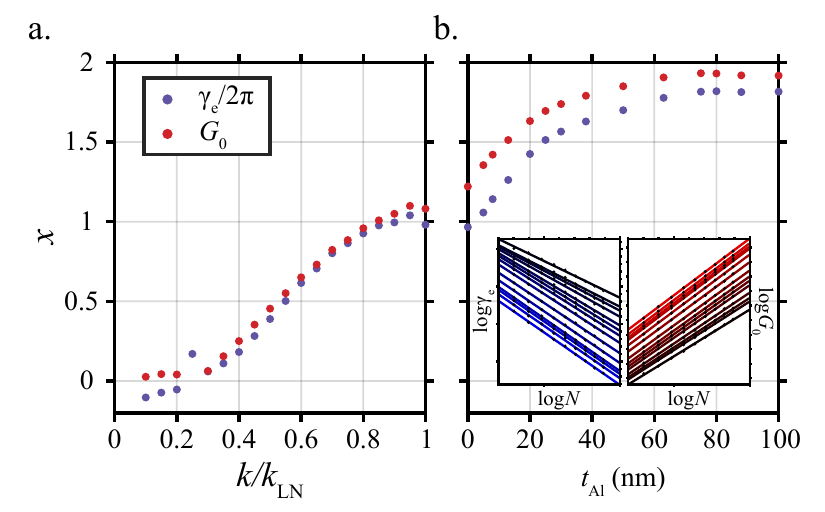}
    \caption{By computing \(G\prens{\omega}\) and fitting the main lobe to a Lorentzian for IDTs with \(N\) ranging from $10$ to $18$, we can extract the exponent \(x\) with which the peak conductance \(G_0\) and bandwidth \(\gamma\) scales.  Sweeping \(k_\trm{eff}^2\) (\textbf{a}) and the electrode thickness \(t_\trm{Al}\) (\textbf{b}) reveals their independent contributions to the resonant behavior of the IDTs.  When \(t_\trm{Al} = 0\) the power-laws strongly deviate from the impulse response model \(x = 0\) due to the large \(k_\trm{eff}^2\) of these IDTs.  Power-law fits to the \(t_\trm{Al}\) sweep are inset to \textbf{b.}.}
    \label{fig:power-law}
\end{figure}

In Figure~\ref{fig:power-law}, it is clear that the exponent \(x\) fit from \(G_0\) matches that of \(\gamma\) such that \(G_0\gamma\) remains proportional to \(N\).  
This shows the trade-off between peak conductance and bandwidth.
The peak conductance is enhanced by the quality factor of the IDT (Appendix~\ref{app:Q}).
Material losses decrease \(Q\) and therefore \(G_0\).
These results are consequences of the net conductance's extensivity and insensitivity to loss on which we will report in a separate manuscript.

\section{Computing $\gamma$}
\label{app:Q}

The bandwidth of a lossless IDT is equal to \(\gamma_\trm{e}\), the rate at which energy leaks out of the transducer. The quality factor is the ratio of the frequency and bandwidth, or equivalently, given by the product of the group delay $\tau_\text{g}=\partial\phi_{12}/\partial \omega$ and frequency:
\begin{align}
	Q &= \frac{\omega_0}{\gamma_\trm{e}}=\frac{\tau_\text{g}\omega_0}{2} \\
	  &= -\frac{1}{2}\max_\omega\left[ \omega \frac{\partial\phi_{21}}{\partial\omega} \right]
	\label{eq:Q}
\end{align}
where \(\phi_{21}\) is the phase difference between the drive \(V(\omega)\) and the emitted mechanical wave \(u_x\).
Reflections increase the delay and therefore the \(Q\) of the IDT.
For these highly resonant IDTs, \(Q\) can also be inferred from the FWHM of a Lorentzian fit to \(G\prens{\omega}\) or from the eigenvalue \(\lambda\) of the resonance
\[Q = \frac{\trm{Re}\lambda}{2\trm{Im}\lambda}.\]

\end{document}